\documentclass[a4paper]{article}
\usepackage{natbib,amsmath,amsfonts,lup,fancyheadings}

\newcommand{\ket}[2]{|#1\rangle _{#2}}

\newcommand{\up}{\!\uparrow}
\newcommand{\down}{\!\downarrow}

\title{Entanglement and Relativity\thanks{Thanks are due to Jon Barrett, Simon Saunders and David Wallace for very useful discussion of various of the topics raised here. C.G.T gratefully acknowledges the support of a studentship from the UK Arts and Humanities Research Board.}}
\author{C.G.\,Timpson and H.R.\,Brown}
\date{3 October 2002}

\bibpunct{[}{]}{;}{a}{}{,}
\begin{document}
\maketitle

\begin{abstract}
This paper surveys some of the questions that arise
when we consider how entanglement and relativity are related via the notion of non-locality.
We begin by reviewing the role of entangled states in Bell inequality violation and question whether the associated notions of non-locality lead to problems with relativity. The use of entanglement and wavefunction collapse in Einstein's famous incompleteness argument is then considered, before we go on to see how the issue of non-locality is transformed if one considers quantum mechanics without collapse to be a complete theory, as in the Everett interpretation. 
\end{abstract}

\section{Introduction}
Entanglement is a property of certain quantum states, relativity a constraint on acceptable physical theories (perhaps also a `theory of spacetime'); relativistic versions of quantum theory exist. Why need we say more?

The immediate response is that entanglement and relativity seem to run up against one another in the spectre of non-locality. We face the question of whether the special sorts of correlation that quantum mechanics allows between spatially separated systems in entangled states, correlations often simply dubbed `non-local', can be consistent with the strictures of relativity\footnote{We shall not be concerned in this paper with the further interesting question of how entanglement transforms under Lorentz boosts; see \citet{peresqmentropysr,gingrich} for investigation of this topic.}. We should begin, however, by noting that the relations of both entanglement and relativity to non-locality are rather subtle, and that the notion of non-locality itself is rather vague, as we shall presently see.
    
The interesting properties of entangled states were emphasised by Einstein, Podolsky and Rosen (EPR) in their famous 1935 paper\nocite{EPR}; the term `entanglement' itself was coined by Schr\"odinger in his work on the quantum correlations that was stimulated by EPR \citep{schrodinger1,schrodinger2}. With the rapid development of quantum information theory over the last 10-15 years and the recognition that entanglement can function as a communication resource, great strides have been made in understanding the properties of entangled states, in particular, with the development of quantitative theories of bipartite (two party) entanglement and the recognition of qualitatively distinct forms of entanglement. (We will begin by  considering mainly the better understood bipartite entanglement). Entanglement assisted communication, an important novel aspect of quantum information theory, also provides a new context in which the relations between entanglement, non-locality and relativity may be explored.

In the following section, the question of how entanglement is related to the notions of non-locality arising from the work of Bell is reviewed, and how this impinges on the constraints of relativity considered. Section~\ref{einstein relativity separability} is concerned with the twin use of entanglement and wavefunction collapse in Einstein's incompleteness argument, recalling the interesting fact that Einstein's true concern in the EPR argument appears not to have been with relativity.
It is often asserted that the Everett interpretation provides an understanding of quantum mechanics unblemished by any taint of non-locality; this claim is assessed in Section~\ref{everett} and the opportunity taken to examine whether a novel sort of non-locality apparently associated with the use of entanglement in dense coding and teleportation might give cause for a new concern.

\section{Entanglement, Non-Locality and Bell Inequalities}\label{ent non-loc bi}

A state is called entangled if it is not separable, that is, if it cannot be written in the form:
\[ |\Psi\rangle_{AB}=|\phi\rangle_{A}\otimes |\psi\rangle_{B}, \textrm{ for pure, or  } \rho_{AB}=\sum_{i}\alpha_{i}\rho_{A}^{i}\otimes \rho_{B}^{i},\textrm{  for mixed states,}\]
where $\alpha_{i} > 0, \sum_{i}\alpha_{i}=1$ and $A$, $B$ label the two distinct subsystems. The case of pure states of bipartite systems is made particularly simple by the existence of the Schmidt decomposition - such states can always be written in the form:
\begin{equation}\label{schmidt}
|\Psi\rangle_{AB}=\sum_{i}\sqrt{p_{i}}\,|\bar{\phi_{i}}\rangle_{A}\otimes |\bar{\psi_{i}}\rangle_{B},
\end{equation}
where $\{|\bar{\phi_{i}}\rangle\},\{|\bar{\psi_{i}}\rangle\}$ are orthonormal bases for systems $A$ and $B$ respectively, and $p_{i}$ are the (non-zero) eigenvalues of the reduced density matrix of $A$. The number of coefficients in any decomposition of the form (\ref{schmidt}) is fixed for a given state $|\Psi\rangle_{AB}$, hence if a state is separable (unentangled), there is only one term in the Schmidt decomposition, and conversely\footnote{For the mixed state case, this simple test does not exist, but progress has been made in providing necessary and sufficient conditions for entanglement for certain systems e.g. $2\otimes 2$ and $2\otimes 3$ dimensional systems (see \cite{horodeckis} for a review).}. The measure of degree of entanglement is also particularly simple for bipartite pure states, being given by the von Neumann entropy of the reduced density matrix of $A$ or $B$ (these entropies being equal, from (\ref{schmidt})).   

To see something of the relation of entanglement to non-locality, we now need to say a little about non-locality. 
One precise notion of non-locality is due, of course, to Bell. Having noted that the de Broglie-Bohm theory incorporates a mechanism whereby the arrangement of one piece of apparatus may affect the outcomes of distant measurements (due to the interdependence of the spacetime trajectories of particles in entangled states), Bell posed the question of whether peculiar properties of this sort might be true of any attempted hidden variable completion of quantum mechanics \citep{bellhiddenvariable}. He was shortly to show (famously) that this is indeed the case \citep{bell1964}.

Consider two measurements, selected and carried out at spacelike separation on a pair of particles initially prepared in some state and then moved apart. The outcomes of the measurements are denoted by $A$ and $B$, and the settings of the apparatuses by $\mathbf{a}$ and $\mathbf{b}$. (We call this scenario a Bell-type experiment.)
We now imagine adding parameters $\lambda$ to the description of the experiment in such a way that the outcomes of the measurements are fully determined by specification of $\lambda$ and the settings $\mathbf{a}$ and $\mathbf{b}$. We also make an assumption of locality - that the outcome of a particular measurement depends \textit{only} on $\lambda$ and on the setting of the apparatus doing the measuring. That is, the outcomes are represented by functions $A(\mathbf{a},\lambda), B(\mathbf{b}, \lambda)$; the outcome $A$ does not depend on the setting $\mathbf{b}$, nor $B$ on $\mathbf{a}$. The parameter (or `hidden variable') $\lambda$ is taken to be chosen from a space $\Lambda$ with a probability distribution $\rho(\lambda)$ over it, and expectation values for these pairs of measurements will then take the form
\begin{equation}\label{expectation}
E(\mathbf{a},\mathbf{b})=\int \!A(\mathbf{a},\lambda)B(\mathbf{b},\lambda)\rho(\lambda)d\lambda.
\end{equation}

This sort of theory is known as a deterministic hidden variable theory. From the expression (\ref{expectation}), a variety of inequalities (Bell inequalities) for the observable correlations in pairs of measurements follow. Violation of such an inequality implies that the correlations under consideration cannot be explained by a deterministic hidden variable model without denying the locality condition and allowing the setting of one apparatus to affect the outcome obtained by the other. It turns out that the quantum predictions for appropriately chosen measurements on, for example, a singlet state, violate a Bell inequality; and the implication is that any attempt to model the quantum correlations by a deterministic hidden variable theory must invoke some non-local mechanism that allows the setting of an apparatus to affect (instantaneously) the outcome of a distant experiment (analogously to the situation in the de Broglie-Bohm theory). Entanglement is a necessary condition for Bell inequality violation, so it is entanglement that makes quantum mechanics inconsistent with a description in terms of a local deterministic hidden variables theory. This is one sense of non-locality.    

The discussion of Bell inequalities was later generalised to the case of \textit{stochastic} hidden variable theories \citep{ch1974,bell1976}, leading to a somewhat different notion of non-locality. In such a stochastic theory, specification of the variables $\lambda$ only determines the \textit{probabilities} of measurement outcomes.

Bell begins his discussion with an intuitive notion of \textit{local causality}, that events in one spacetime region cannot be causes of events in another, spacelike separated, region. He then goes on to define a model of a correlation experiment as being \textit{locally causal} if the probability distribution for the outcomes of the measurements factorises when conditioned on the `hidden' state $\lambda$ in the overlap of the past light cones of the measurement events. Thus the requirement for a locally causal theory is that \[p_{\,\mathbf{a},\mathbf{b}}(A\wedge B|\lambda)=p_{\,\mathbf{a}}(A|\lambda)p_{\,\mathbf{b}}(B|\lambda),\] where, as before, $A$ and $B$ denote the outcomes of spacelike separated measurements and $\mathbf{a},\mathbf{b}$ the apparatus settings.

Imposing this requirement amounts to saying that once all the possible common causes of the two events are taken into account (which, guided by classical relativistic intuitions, we take to reside in their joint past), we expect the probability distributions for the measurement outcomes to be independent and no longer display any correlations\footnote{The correlation coefficient between two random variables $x$ and $y$ is given by the covariance of $x$ and $y$, $\mathrm{cov}(x,y)=\langle (x-\langle x\rangle)(y-\langle y\rangle)\rangle$, divided by the square root of the product of the variances. If $x$ and $y$ are statistically independent, $p(x\wedge y)=p(x)p(y)$, then $\mathrm{cov}(x,y)=0$ and the variables are uncorrelated.}. Again, from the assumption of factorisability, a number of inequalities can be derived which are violated for some measurements on entangled quantum states\footnote{If we wanted to be more precise, we would need also to assume, for example, that the types of measurement $A$ and $B$ chosen did not depend in a conspiratorial way on $\lambda$.}, leading to the conclusion that such correlations cannot be modelled by a locally causal theory. In fact, (note that we could take $\lambda$ simply to be the quantum state of the joint system) we know that quantum mechanics itself is not a locally causal theory, as it can be seen directly that factorisability will fail for some measurements on entangled states.

However, it is important to note that failure of local causality in Bell's sense does not entail the presence of non-local causes. In arriving at the requirement of factorisability it is necessary to assume something like Reichenbach's principle of the common cause; namely, to assume that if correlations are not due to a direct causal link between two events, then they must be due to common causes, such causes having being identified when conditionalisation of the probability distribution results in statistical independence. (Then, arguing contrapositively, if the correlations can't be due to common causes, we can infer that they must be due to direct causes.) Thus although failure of factorisability may imply that quantum correlations cannot be explained by a common cause, it need not imply that there must therefore be direct (and hence non-local) causal links between spacelike separated events: it could be the principle of the common cause that fails. Perhaps it is simply not the case that in quantum mechanics, correlations are always apt for causal explanation\footnote{\label{dickson}In contrast to the present approach, \citet[Chpt. 4]{Maudlin:non-loc} insists that failure of factorisability does imply non-local causation as he adheres to the common cause principle, taking it be almost a tautology that lawlike prediction of correlations is indicative of a causal connection, either directly or via a common cause. One might be worried, however, that this sort of argument does not provide a sufficiently robust notion of cause to imply a genuine notion of non-local action-at-a-distance. \citet{dickson} argues that in the absence of full dynamical specification of a model of Bell experiments (of the sort that the de Broglie-Bohm theory, for instance, provides), as opposed to the rather schematic hidden variable schemes we have been considering, it is in any case precipitate to reach much of a conclusion about whether genuine non-locality is involved in a model.}. When discussing non-locality in the context of Bell inequality violation, then, we see that it is important to distinguish between the non-locality implied for a deterministic hidden variable model of an experiment and a violation of local causality, which latter need not, on its own, imply any non-locality.  

Some terminology: Following \citet{jarrett}, Bell's condition of factorisability is usually analysed as the conjunction of two further conditions, sometimes called \textit{parameter independence} and \textit{outcome independence}\footnote{\citet[pp.94-5]{maudlin} points out that Jarrett's decomposition is not unique; it may nonetheless be a useful distinction; see \citet[\S 6.2.3]{dickson} for further discussion.}. Parameter independence states that the probability distribution for local measurements, conditioned on the hidden variable, should not depend on the setting of a distant measuring apparatus; it is a consequence of the \textit{no-signalling} theorem (which we shall discuss further in the next section) that quantum mechanics, considered as a stochastic hidden variable theory, satisfies this condition. The condition of outcome independence, on the other hand, states that when we have taken into account the settings of the apparatuses on both sides of the experiment and all the relevant factors in the joint past, the probability for an outcome on one side of the experiment should not depend on the actual result at the other. It is this condition that is typically taken to be violated in orthodox quantum mechanics. We shall discuss outcome independence further in the context of the Everett interpretation below (Section~\ref{everett}); for a discussion of the relation between outcome independence and Reichenbach's principle of the common cause, see \citet{harvey:pass}. 

Previously we said that entanglement was a necessary condition for Bell inequality violation; somewhat surprisingly, it turns out that the converse is not true. \citet{werner} showed that a local hidden variable model could be constructed for a certain class of entangled mixed states (`Werner states'), which means that entanglement is not a sufficient condition for non-locality in the sense of Bell inequality violation; the two terms are not synonymous (although it was subsequently established that entanglement does always imply violation of some Bell inequality for the restricted case of pure states \citep{peresgisin,popescurohrlich}). The plot was further thickened when Popescu demonstrated that the Werner states for dimensions greater than or equal to five could, however, be made to violate a Bell inequality if sequential measurements are allowed, rather than the single measurements of the standard Bell inequality scenario; this property he termed `hidden non-locality' \citep{popescu} (see also \citet{jon1,jon2} for a useful discussion and a further development).    
We see that the relationship between entanglement and Bell inequality violation is indeed subtle, just as the links between Bell inequality violation and non-locality are complex. What, then, of relativistic concerns?

Normally one would vaguely assent to the idea that special relativity implies that the speed of light is a limiting speed, in particular, the limiting speed for causal processes or for signalling. It has long been recognised, however, that the fundamental role of the speed of light in relativity is as the invariant speed of the theory, not, in the first instance, as a limiting speed\footnote{This goes against the `absolute' approach of Robb, Zeeman and others. See e.g. \citet[Chpt.3]{LucasandHodgson}.}. That being said, though, one does still need to pay some attention to the question of superluminal signalling.

The immediate concern with superluminal signals in a relativistic context is familiar - in at least some frames, a superluminal signal is received before it is emitted, so we could arrange for a signal to be sent back that would result in the original signal not being transmitted, hence paradox. Clearly, such signal loops must be impossible. In his influential discussion, though, \citet{Maudlin:non-loc} draws a distinction between superluminal signals \textit{simpliciter} and superluminal signals that allow loops, arguing that it is only the latter that need give rise to inconsistencies with relativity. In our opinion, however, it is not clear that this distinction is actually based on physically reasonable possibilities, which leads us to conclude more simply that superluminal signals cannot satisfactorily be incorporated into Lorentz covariant theories as they would naturally give rise to the possibility of loop paradoxes. 

The fundamental relativistic constraint is that of Lorentz covariance. From the principle of Relativity (roughly, the laws governing the change of physical systems do not depend on which inertial frames those changes are referred to), the Light Postulate (the two-way speed of light in the `resting' frame is independent of the velocity of its source and is isotropic) and some other natural constraints (linearity, the homogeneity of space and time and the isotropy of space), \citet{Einstein1905} derived the appropriate transformations - the Lorentz transformations - between inertial frames. Laws that are not Lorentz covariant would violate the principle of Relativity (given the light postulate), for the laws would depend on which inertial frame they were referred to, which is to say, they would pick out a preferred frame of reference.

How, then, do the notions of non-locality that we have seen to be associated with some forms of entanglement in the violation of Bell inequalities relate to relativity? Only indirectly, at best. Violation of Bell's notion of local causality does not on its own imply non-local causation, hence there is no immediate suggestion of any possible conflict with Lorentz covariance; and although the implied non-local mechanism in a deterministic hidden variable theory would pick out a preferred inertial frame and violate Lorentz covariance\footnote{At least when a full dynamical specification of the theory was forthcoming; cf. the comment on Dickson, footnote \ref{dickson}.}, this would only be of any significance if we were actually to choose to adopt such a model\footnote{Even then, surprisingly, one might still argue that there is no real violation of relativity; some have suggested that it is only \textit{observable}, or empirically accessible quantities that need be Lorentz invariant \citep[cf.][]{brownelbyweingard:1996}. This would imply that even the de Broglie-Bohm theory might be consistent with relativity, despite its manifest non-locality.}.     

A more direct link between entanglement and relativity would be in the offing, however, if we were to take collapse of the wavefunction seriously, a matter we have not so far addressed. Collapse seems just the sort of process of action-at-a-distance that would be inimical to relativity, and in the context of EPR or Bell-type experiments, the processes of collapse that would be needed to explain the observed correlations raise profound difficulties with the requirement of Lorentz covariance. Some Lorentz covariant theories of dynamical collapse have been proposed (e.g. \citet{fleming,ghiradi}), but rather than consider these, our attention now turns instead to consider how relativity and collapse figure in the context of Einstein's celebrated discussions of the interpretation of quantum theory, before we go on to see how the issue of non-locality is transformed if one denies collapse, as in the Everett theory. 

\section{Einstein, Relativity and Separability}\label{einstein relativity separability}

Einstein's scepticism about quantum mechanics, or at least his opposition to the quantum theory as preached by the Copenhagen school, is well known. As the father of relativity and in the context of the EPR paper, it is tempting to see this opposition as based on his recognising that the sort of correlations allowed by entangled states will conflict with relativity at some level. At the 1927 Solvay Conference, Einstein did indeed convict quantum theory, considered as a complete theory of individual processes, on the grounds of conflict with relativity. At this early stage, though, his concern was with the action-at-a-distance implied by the collapse on measurement of the wavefunction of a single particle diffracted at a slit (a process that somehow stops the spatially extended wavefunction from producing an effect at two or more places on the detecting screen), rather than anything to do with entanglement.

In fact, as several commentators have remarked, Einstein's fundamental worry in the EPR paper, a worry more faithfully expressed in his later expositions of the argument, was not directly to do with relativity. Rather, his opposition to quantum theory was based on the fact that, if considered complete, the theory violates a principle of separability for spatially separated systems. What he had in mind is expressed in the following passage:
\begin{quotation}\small
If one asks what, irrespective of quantum mechanics, is characteristic of the world of ideas of physics, one is first of all struck by the following: the concepts of physics relate to a real outside world, that is, ideas are established relating to things such as bodies, fields, etc., which claim `real existence' that is independent of the perceiving subject...It is further characteristic of these physical objects that they are thought of as arranged in a space-time continuum. An essential aspect of this arrangement of things in physics is that they lay claim, at a certain time, to an existence independent of one another, provided these objects `are situated in different parts of space'. Unless one makes this kind of assumption about the independence of the existence (the `being thus') of objects which are far apart from one another in space - which stems in the first place from everyday thinking - physical thinking in the familiar sense would not be possible. It is also hard to see any way of formulating and testing the laws of physics unless one makes a clear distinction of this kind.

...The following idea characterises the relative independence of objects far apart in space (A and B): external influence on A has no direct influence on B; this is known as the `principle of contiguity', which is used consistently only in the field theory. If this axiom were to be completely abolished, the idea of the existence of (quasi-)enclosed systems, and thereby the postulation of laws which can be checked empirically in the accepted sense, would become impossible. \citep{einstein}
\end{quotation}     

In this passage, Einstein's description of the grounds for a principle of separability takes on something of the form of a transcendental argument - separability is presented as a condition on the very possibility of framing empirical laws.   
From the EPR argument, however, it follows that separability is not consistent with the thought that quantum mechanics is complete.

If we consider two entangled systems, $A$ and $B$, in a pure state, the type of measurement made on $A$ will determine the (pure) state that is ascribed to $B$, which may be at spacelike separation. Any number of different measurements could be performed on $A$, each of which would imply a different final state for $B$. From the principle of separability, however, the real state of a distant system cannot be affected by the type of measurement performed locally or indeed on whether any measurement is performed at all; separability then requires that the real state of $B$ (the physically real in the region of space occupied by $B$) has \textit{all} of these different possible quantum states associated with it simultaneously, a conclusion clearly inconsistent with the wavefunction being a \textit{complete} description\footnote{This is the simplified version of the EPR incompletness argument which Einstein preferred. It is interesting to note the dialectical significance of Einstein's use of entangled systems in the argument for incompleteness. His earlier attempt to argue for incompleteness could be blocked by Bohr's manouevre of invoking the unavoidable disturbance resulting from measurement in order to explain non-classical behaviour and the appearance of probabilities in the quantum description of experiments. Using the correlations implicit in entangled systems to prepare a state cleverly circumvents the disturbance doctrine, as there is no possibility of the mechanical disturbance Bohr had in mind being involved \citep[cf.][]{fine}.}. 

Since the principle of separability is supported by his quasi-transcendental argument, Einstein would seem to be on firm ground in denying that quantum mechanics can be a complete theory. On further consideration, however, the argument for separability can begin to look a little shaky. Isn't quantum mechanics itself a successful empirical theory? Is it not then simply a counterexample to the suggestion that physical theory would be impossible in the presence of non-separability? As Maudlin puts it: `quantum mechanics has been precisely formulated and rigorously tested, so if it indeed fails to display the structure Einstein describes, it also immediately refutes his worries' \citep[p.49]{maudlin}.

As implied in Maudlin's phrasing (note the conditional), some care is required with this response. Since it is precisely the question of how the formalism of quantum mechanics is to be interpreted that is at issue, we are always free to ask \textit{what it is} that is supposed to have been tested by the empirical success of the predictions of quantum theory. A complete and non-separable theory, or an incomplete and statistical one? Appeal to bare empirical success appears  insufficient to decide the question of the meaning of the formalism. Let us, then, try to sharpen the intuition that orthodox quantum theory may constitute a counterexample to Einstein's argument.

It is useful to look in a little more detail at the principle of separability. Although it is not clear that he does so in the above quoted passage, elsewhere at least, Einstein distinguished between what might be called separability proper and locality (\citet[p.164]{borneinstein};\citet[p.85]{einsteinauto}; \citep[cf.][]{howard}). Separability proper is the requirement that separated objects have their own independent real states (in order that physics can have a subject matter, the world be divided up into pieces about which statements can be made); locality is the requirement that the real state of one system remain unaffected by changes to a distant system.
One would usually take this locality condition to fail in orthodox quantum mechanics with collapse\footnote{Some commentators have suggested that Bohr for one, however, upheld locality at the expense of separability proper, see e.g. \citep[\S 8.7]{murdoch}.}. The justification for the feeling that the orthodox theory provides a counterexample to Einstein's transcendental argument is that this failure of locality is relatively benign; it does not seem to make the testing of predictions for isolated systems impossible. It is important to note why. The predictions made by quantum mechanics are of the probabilities for the outcomes of measurements. It is established by the no-signalling theorem, however, that the \textit{probabilities} for the outcomes of any measurement on a given sub-system, as opposed to the state of that system, cannot be affected by operations performed on a distant system, even in the presence of entanglement. Thus the no-signalling theorem entails that quantum theory would remain empirically testable, despite violating locality\footnote{An early version of the no-signalling theorem, specialised to the case of spin 1/2 EPR-type experiments appears in \citet{bohm}. Later, more general versions are given by \citet{tausk,eberhard,grw}. See also \citet{shimony,redhead}.}.

There might remain the worry from the Einsteinian point of view that testability at the level of statistical predictions does not help with the problem that the proper descriptive task of a theory may be rendered difficult given a violation of locality. Local experiments may have difficulty in determining the real states, if such there be; but this is a different issue from the question of whether it is possible to state empirical laws. Consider, for example, the de Broglie-Bohm theory. Here we have perfectly coherently stated dynamical laws, yet if the distribution for particle co-ordinates is given (as usual) by the modulus squared of the wavefunction, it is impossible to know their positions. So it does seem that counterexamples can be given to Einstein's transcendental argument, orthodox quantum theory with collapse being one of them. With this in mind, it is somewhat reassuring that having presented the transcendental argument, Einstein went on to reach a more conciliatory conclusion:
\begin{quote} 
There seems to me no doubt that those physicists who regard the descriptive methods of quantum mechanics as definitive in principle would...drop the requirement...for the independent existence of the physical reality present in different parts of space; they would be justified in pointing out that quantum theory nowhere makes explicit use of this requirement.
I admit this, but would point out: when I consider the physical phenomena known to me...I still cannot find any fact anywhere which would make it appear likely that [the requirement] will have to be abandoned.
\cite[p.172]{einstein} 
\end{quote}

Having introduced the no-signalling theorem, we should close this section by remarking that this theorem is of crucial importance in saving quantum mechanics from explicit non-locality; moreover, a form of non-locality that would lead to direct conflict with relativity\footnote{The no-signalling theorem rules out the possibility of signalling using entangled systems, but note that in the context of non-relativistic quantum mechanics, signalling at arbitrary speeds is nonetheless possible by other means. For example, if the walls are removed from a box in which a particle has been confined, then,  instantaneously, there is a non-zero probability of the particle being found in \textit{any} region of space.}. For if the probability distribution for measurement outcomes on a distant system could be affected locally, this would provide the basis for superluminal signalling processes that could lead to the possibility of temporal loop paradoxes; and such processes, we have suggested, cannot be incorporated into a Lorentz covariant dynamical theory.

\section{Non-locality in the Everett interpretation}\label{everett}

Einstein's argument for the incompleteness of quantum mechanics made crucial use of the notion of collapse along with that of entanglement. It is interesting to see what happens if one attempts to treat quantum theory \textit{without} collapse as a complete theory. This way lies the Everett approach, to which we shall now turn. 

It is a commonplace that the Everett interpretation\footnote{It is perhaps worth noting that there have been a number of different attempts to develop Everett's original ideas into a full-blown interpretation of quantum theory (Many Minds, Many Worlds...). The most satisfactory of these, in our view, is the `relativism' of \citet{simon1,simon2,simon3,simonrelativism} (see also \citep{wallace}) which resolves the preferred basis problem and has made considerable progress on the question of the meaning of probability in Everett, these two being the main problems that have threatened the coherence of the approach. It is this version that should be considered, then, when a detailed ontological picture is desired in what follows, although nothing much in the discussion of non-locality should turn on this.} provides us with a picture in which non-locality plays no part; indeed, this is often presented as one of the main selling points of the approach. Everett himself stated rather dismissively that his relative state interpretation clarified the `fictitious paradox' of Einstein, Podolsky and Rosen \citep[\S 5.3]{everett}; \citet{page} illustrates in some detail how the absence of collapse in the Everett picture allows one to circumvent the argument for incompleteness.

Two things are important for the apparent avoidance of non-locality in Everett. The first is that we are dealing with a no-collapse theory, a theory of the universal wavefunction in which there is only ever unitary evolution. Removing collapse, with its peculiar character of action-at-a-distance, we immediately do away with one obvious source of non-locality. The second, crucially important, factor (compare the de Broglie-Bohm theory) is that in the Everett approach, the result of a measurement is not the obtaining of one definite value of an observable at the expense of other possible values.

In the presence of entanglement, sub-systems of a joint system typically do not possess their own state (i.e. are not in an eigenstate of any observable), but only a reduced density matrix; it is the system as a whole which alone has a definite state. In the Everett approach, however, what claim importance are the states relative to other states in an expansion of the wavefunction. A given sub-system might not, then, be in any definite state on its own, but relative to some arbitrarily chosen state of another sub-system, it \textit{will} be in an eigenstate of an observable. That is, it possesses a definite value of the observable relative to the chosen state of the other system. This allows us to give an explanation of what happens on measurement. Measurement interactions, on the Everett picture, are simply (unitary!) interactions which have been chosen so as to correlate states of the system being measured to states of a measuring apparatus. Following an ideal first-kind (non-disturbing) measurement, the measured system will be in a definite state (eigenstate of the measured observable) relative to the indicator states of the measuring apparatus and ultimately, relative to an observer. 
If there are a number of different possible outcomes for a given measurement, a state corresponding to each outcome will have become definite relative to \textit{different} apparatus (observing) states, following the interaction. Thus the result of a measurement is not that one definite value alone from the range of possible values of an observable obtains, but that each outcome becomes definite relative to a different state of the observer or apparatus\footnote{The situation becomes more complicated when we consider the more physically realistic case of measurements which are not of the first kind; in some cases, for example, the object system may even be destroyed in the process of measurement. What is important for a measurement to have taken place is that measuring apparatus and object system were coupled together in such a way that if the object system had been in an eigenstate of the observable being measured prior to measurement, then the subsequent state of the measuring apparatus would allow us to infer what that eigenstate was. In this more general framework the importance is not so much that the object system is left in a eigenstate of the observable relative to the indicator state of the measuring apparatus, but that we have definite indicator states relative to macroscopic observables. }.

Given this sort of account of measurement, there appears to be no non-locality, and certainly no conflict with relativity in Everett. The obtaining of definite values of an observable is just the process of one system coming to have a certain state relative to another, a result of local interactions governed by the unitary dynamics. Furthermore, when considering spacelike separated measurements on an entangled system, say measurements of parallel components of spin on systems in a singlet state, there is no question of the obtaining of a determinate value for one sub-system requiring that the distant system acquire the corresponding determinate value, instead of another. Both sets of anti-correlated values are realised (become definite) relative to different observing states; there is, as it were, no dash to ensure agreement between the two sides to be a source of non-locality and potentially give rise to problems with Lorentz covariance.

Let us consider the example of the Bohm version of the EPR experiment in a little more detail. We begin with two spin-1/2 systems (labelled 1 and 2) prepared in a singlet state. The relevant degrees of freedom of measuring apparatuses $m_{A}, m_{B}$ in regions $A$ and $B$ of space we can model as two-state systems with basis states $\{\ket{\up}{A},\ket{\down}{A}\}$,\linebreak $\{\ket{\up}{B},\ket{\down}{B}\}$ respectively. (In what follows we shall suppress the explicit tensor product sign). Measurement by apparatus $A$ of the component of spin at an angle $\theta$ from the $z$-axis on system 1 would have the effect
\begin{equation}\label{measurement}
U({\theta}) \left\{ \begin{array}{ccc}
			\ket{\up_{\theta}}{1}\ket{\up_{\theta}}{A} & \mapsto & \ket{\up_{\theta}}{1}\ket{\up_{\theta}}{A} \\
			\ket{\down_{\theta}}{1}\ket{\up_{\theta}}{A} & \mapsto & \ket{\down_{\theta}}{1}\ket{\down_{\theta}}{A}      \end{array}
\right.,
\end{equation}
where $\ket{\up_{\theta}}{},\ket{\down_{\theta}}{}$ are eigenstates of spin in the rotated direction; similarly for measurement of system 2 by apparatus $B$.

If we treat the case of parallel spin measurements at $A$ and $B$ first, the initial state of the whole system is

\begin{equation}\label{parallel initial}
\ket{\up}{A}\tfrac{1}{\sqrt{2}}\bigl(\ket{\up}{1}\ket{\down}{2} - \ket{\down}{1}\ket{\up}{2}\bigr) \ket{\up}{B}.
\end{equation}
Note that the states of the measuring apparatuses factorise; at this stage, they are independent of the states of the spin-1/2 particles. For the measurements to be made, systems 1 and 2 are taken to regions $A$ and $B$ respectively and measurement interactions of the form (\ref{measurement}) occur (the time order of these interactions is immaterial); we finish with the state

\begin{equation}\label{parallel final}
\tfrac{1}{\sqrt{2}}\bigl( \ket{\up}{A}\ket{\up}{1}\ket{\down}{2}\ket{\down}{B}-\ket{\down}{A}\ket{\down}{1}\ket{\up}{2}\ket{\up}{B}\bigr).
\end{equation}

It is important to note that following the measurements at $A$ and $B$, not only does each measured system have a definite spin state relative to the indicator state of the device that has measured it, but the systems and measuring apparatuses in each region (e.g. system 1 and apparatus $m_{A}$ in $A$) have definite spin and indicator states relative to definite spin and indicator states of the system and apparatus in the \textit{other} region (e.g. 2,$m_{B}$ in $B$). That is, following the two local measurements, from the point of view of the systems in one region, the states of the systems in the far region correspond to a definite, in fact perfectly anti-correlated, measurement outcome. This is in contrast to the general case of non-parallel spin measurements at $A$ and $B$.
Here we may write the initial state as

\begin{equation}\label{non-parallel initial 1}
\ket{\up}{A}\tfrac{1}{\sqrt{2}}\bigl( \ket{\up}{1}\ket{\down}{2} - \ket{\down}{1}\ket{\up}{2}\bigr) \ket{\up_{\theta}}{B},
\end{equation}
where we have assumed a relative angle $\theta$ between the directions of measurement. If we write $\ket{\up}{}=\alpha \ket{\up_{\theta}}{}\! + \!\beta \ket{\down_{\theta}}{},\; \ket{\down}{}=\alpha^{\prime}\ket{\up_{\theta}}{}\! + \!\beta^{\prime}\ket{\down_{\theta}}{}$, this becomes

\begin{equation}\label{non-parallel initial 2}
\ket{\up}{A}\tfrac{1}{\sqrt{2}}\Bigl[\, \ket{\up}{1} \bigl( \alpha \ket{\up_{\theta}}{2}\! +\! \beta \ket{\down_{\theta}}{2}\bigr) - \ket{\down}{1} \bigl( \alpha^{\prime}\ket{\up_{\theta}}{2}\! +\! \beta^{\prime}\ket{\down_{\theta}}{2} \bigr)\Bigr]\: \ket{\up_{\theta}}{B}.
\end{equation}
We can then see that following the measurements at $A$ and $B$, we will have

\begin{equation}\label{non-parallel middle}
\begin{split}
\frac{1}{\sqrt{2}} \Bigl[ \,\ket{\up}{A}\ket{\up}{1}\:\bigl(\alpha \ket{\up_{\theta}}{2}\ket{\up_{\theta}}{B}& + \beta \ket{\down_{\theta}}{2}\ket{\down_{\theta}}{B}\bigr)   \\
	 - \,&\ket{\down}{A}\ket{\down}{1}\,\bigl(\alpha^{\prime}\ket{\up_{\theta}}{2}\ket{\up_{\theta}}{B} + \beta^{\prime}\ket{\down_{\theta}}{2}\ket{\down_{\theta}}{B}\bigr)\Bigr].  
\end{split}
\end{equation} 
Here, relative to states representing a definite outcome of measurement in region $A$, there is no definite outcome in $B$; system 2 and apparatus $m_{B}$ are just entangled, with no definite spin and indicator states. Similarly, from the point of view of definite spin and indicator states of 2 and $m_{B}$ (a definite outcome in region $B$), there is no definite outcome of measurement in region $A$.

For non-parallel spin measurements, then, unlike the parallel case, we need to perform a third measurement, comparing the outcomes from $A$ and $B$, in order to make definite spin and indicator states from one side definite relative to definite spin and indicator states from the other. Thus systems from $A$ and $B$ have to be brought together and a joint measurement performed, leading to a state like:

\begin{multline}\label{non-parallel final}
\frac{1}{\sqrt{2}} \Bigl[\, \alpha \ket{\up}{A}\ket{\up}{1}\ket{\up_{\theta}}{2}\ket{\up_{\theta}}{B}\ket{\up\up}{C} + \beta \ket{\up}{A}\ket{\up}{1}\ket{\down_{\theta}}{2}\ket{\down_{\theta}}{B}\ket{\up\down}{C}  \\
 - \alpha^{\prime}\ket{\down}{A}\ket{\down}{1}\ket{\up_{\theta}}{2}\ket{\up_{\theta}}{B}\ket{\down\up}{C} - \beta^{\prime} \ket{\down}{A}\ket{\down}{1}\ket{\down_{\theta}}{2}\ket{\down_{\theta}}{B}\ket{\down\down}{C}\Bigr],  
\end{multline}     
where the states $\ket{\up\up}{C}$ etc. are the indicator states of the comparing apparatus. Following this third measurement, which can only take place in the overlap of the future light cones of the measurements at $A$ and $B$, a definite outcome for the spin measurement in one region finally obtains, relative to a definite outcome for the measurement in the other. We note that it is only with a joint measurement of this sort that the non-factorisable joint probability distribution for the outcomes of the non-parallel spin measurements is observable. Indeed, in the Everett picture, prior to this third stage of measurement when systems from $A$ and $B$ have been brought together, the  joint probability distribution for the outcomes of the spin measurements can only denote what we might call a joint propensity, namely a propensity of the two systems to display the appropriate probability distribution \textit{if} brought together \textit{and} the comparison measurement performed.

The distinction introduced in Section~\ref{ent non-loc bi} between parameter independence and outcome independence in the framework of stochastic hidden variable theories provides a common terminology in which the oddity of the quantum correlations inherent in entangled states is discussed. As we mentioned, the no-signalling theorem ensures that parameter independence is satisfied for orthodox quantum mechanics and it is violation of outcome independence that one usually takes to be responsible for failure of factorisability. Recall that the condition of outcome independence states that once one has taken into account the settings of the apparatuses on both sides of the experiment and all the relevant factors in the joint past, the probability for an outcome on one side of the experiment should not depend on the actual result at the other. Violation of this condition, then, presents us with a very odd situation. \textit{For how could it be true of two stochastic processes which are distinct and supposed to be \emph{irreducibly random} that their outcomes may nonetheless display correlations}?
(One might say that this question captures the essence of the concerns raised by the EPR argument, but formulated in a way that requires neither explicit appeal to collapse nor to perfect correlations (as in the case of parallel spin measurements on the singlet state).)

The Everett interpretation can help us with the oddity of the violation of outcome independence and with understanding how the quantum correlations come about in general. The important thing to note is that the obtaining of the relevant correlations in the Everett picture is not the result of a stochastic process, but of a deterministic one.
Given the initial state of all the systems, the appropriate correlations follow deterministically, given the interactions that the systems undergo. Correlations between measurement outcomes obtain if definite post-measurement indicator states of the various measuring apparatuses are definite relative to one another.
If we have a complete, deterministic story about how \textit{this} can come about, as we do, then it is difficult to see what more could be demanded in explanation of how the quantum correlations come about. (Of course, the general question of understanding the nature and capacities of the correlations inherent in entangled states remains, but as a separate issue.) 

We have seen how the story goes in the EPR-type scenario: for parallel  measurements, given the initial entangled singlet state and given the measurement interactions, it was always going to be the case that an `up' outcome on side $A$ would be correlated with a `down' outcome on side $B$ and a `down' at $A$ with an `up' at $B$ (both outcomes co-existing in the Everett sense). For the case of non-parallel measurements, correlations don't immediately obtain after the two spin measurements, but again, it follows deterministically that the desired correlations obtain if the necessary third measurement is later performed. 
With regard to the question of outcome independence, we see that we avoid the tricky problem of having to explain how the condition is violated, as from the point of view of Everett, the scenario is not one of distinct stochastic processes mysteriously producing correlated results, but of correlations arising deterministically from an initial entangled state. 

It is interesting to compare how Everett deals with the apparent non-locality and potential problems with Lorentz covariance that arise with entanglement and how a statistical interpretation does so. (By statistical interpretation, we mean an interpretation in which quantum mechanics merely describes probabilities for measurement outcomes for ensembles, there is no description of individual systems and collapse does not correspond to any real physical process for individual systems). 
In neither case do we have collapse, nor anything added to the quantum formalism, hence in neither case do we face a problem of the form: how can there be correlations between space-like separated events that cannot be explained by factors in their joint past? The only question that can remain is whether any problems arise from the fact that things can sometimes depend in an interesting way on parameters chosen in space-like separated regions (for example, the non-factorisable probability distributions in EPR-type experiments depending on choices of measurement directions).

Noting that in a relativistic quantum mechanics, the state will behave correctly under Lorentz transformations, it is clear that the no-signalling theorem does all the work necessary to ensure that there will be no problems with relativity on either interpretation. The relevant dependence on the parameters is only in joint probability distributions and these can only be observed at a point, when the appropriate systems have been brought together\footnote{The moral here seems to be: what the tensor product structure of the state space gives with one hand (correlations between separated systems), it takes away with the other (have to perform a joint measurement). From this point of view it appears that there is no \textit{conspiracy} about why quantum mechanics is compatible with relativity; the appropriate thing to say seems to be that quantum mechanics is simply blind to spacetime considerations.}. As for possible non-locality, things are a trifle more involved. If we take the possible effects of measurements to be our main object of concern, one can justifiably say that in both Everett and the statistical interpretation there is no action-at-a-distance, although for slightly different reasons in the two cases. In the statistical interpretation, there is simply no action on measurement, as we are not trying to describe the behaviour of individual systems; in Everett there \textit{is} action on measurement (correlation between systems and apparatus), but it doesn't operate at a distance. 

This latter claim stands in need of some clarification. We emphasized earlier that there is no action-at-a-distance analogous to the effect of collapse in Everett, due to the co-existence of the different measurement outcomes; however it might be argued that this is not quite sufficient to establish that the effect of a measurement is entirely local. Of course, the no-signalling theorem again plays an important role in establishing that local expectation values will be unaffected by a distant measurement, but our worry might be of a different sort. As has become clear from the phenomena of entanglement assisted communication, in the presence of entanglement, \textit{local} unitary operations (of which measurement, of course, is one) can have a significant effect on the \textit{global} state.  

This is most obvious in superdense coding, the first example of entanglement assisted communication \citep{superdense}. Here we imagine two parties, Alice and Bob, who share a maximally entangled state, for example the singlet state of two two-state systems. By applying one of the four operators $\{{\mathbf 1},\sigma_{x},\sigma_{y},\sigma_{z}\}$ to her half of the entangled pair and then sending it to Bob, Alice can succeed in sending him two bits of information rather than the single bit which is the most that one normally expects to be able to encode into a single two-state system. The trick is that applying one of the Pauli operators to Alice's system flips the joint state into one of the others of the four maximally entangled `Bell' states (Table~\ref{bellstates}).
\begin{table}
\begin{center}
\begin{minipage}{2.3in}
\[ \left. \begin{array}{c}
\ket{\phi^{+}}{}=1/\sqrt{2}(\ket{\up}{}\ket{\up}{}+\ket{\down}{}\ket{\down}{})\\
\ket{\phi^{-}}{}=1/\sqrt{2}(\ket{\up}{}\ket{\up}{}-\ket{\down}{}\ket{\down}{})\\
\ket{\psi^{+}}{}=1/\sqrt{2}(\ket{\up}{}\ket{\down}{}+\ket{\down}{}\ket{\up}{})\\
\ket{\psi^{-}}{}=1/\sqrt{2}(\ket{\up}{}\ket{\down}{}-\ket{\down}{}\ket{\up}{})
\end{array} \right\}\]
\end{minipage} \ =  \
\begin{minipage}{1.5in}
\[ \left\{\begin{array}{r}
-i\sigma_{y}\otimes \mathbf{1}\ket{\psi^{-}}{}\\
-\sigma_{x}\otimes \mathbf{1}\ket{\psi^{-}}{}\\
\sigma_{z}\otimes \mathbf{1}\ket{\psi^{-}}{}\\
\mathbf{1}\otimes\mathbf{1}\ket{\psi^{-}}{}
\end{array} \right. \]
\end{minipage}
\end{center}
\caption{The four Bell states, a maximally entangled basis for $2\otimes2$ dim. systems.}\label{bellstates}
\end{table}
When both systems are in his hands, Bob simply needs to perform a Bell basis measurement, whose outcome will then tell him which operator was applied by Alice. In this protocol, the possibility of changing the global state by a local operation is being used to send information in an unexpected way. (\citet{braunstein:twist} argue that the ability to span the set of maximally entangled states by local unitary operators also lies at the heart of \textit{teleportation}, a phenomenon to which we shall shortly turn.)

This sort of global effect of a local operation is certainly striking\footnote{\citet{ekertjozsa} suggest that effects of this sort are also responsible for the exponential speed up achievable by quantum computers for certain computational tasks.}, but it is not clear that it is appropriately called a \textit{non-local} effect. After all, one might argue, the global state is not obviously a spatio-temporal entity, even if we are being robustly realist about it; and unlike the sort of change associated with collapse, it is certainly not the case that any local and non-relational properties of the separated systems change when the operation is performed (i.e. locally observable probability distributions are unchanged).   

How about the other example of entanglement assisted communication, teleportation \citep{teleportation}? Here Alice and Bob again share a maximally entangled state, but this time, Alice is trying to send an unknown quantum state rather than classical information. She begins by performing a Bell basis measurement on her half of the entangled pair and the system whose state she is trying to transmit. This has the effect of flipping Bob's half of the entangled pair into a state which differs from the state Alice is sending by one of four unitary operations, depending on what the outcome of Alice's measurement was. If the output of Alice's measurement is sent to Bob, he can perform the required operation to obtain a system in the state Alice was trying to send. The significant feature of this protocol for our current purposes is that from the Everett perspective, immediately following Alice's measurement and before she sends a record of her outcome to Bob, Bob's system will already have acquired a definite state related to the state Alice is sending, relative to the outcome of Alice's measurement. And this looks like a form of non-locality: the pertinent relative state of Bob's system has come to depend on the parameters characterising the state being sent by Alice, merely as a result of a local operation (measurement) carried out at a distance by Alice, and without any direct interaction between the two sides of the experiment.    

It seems that this appearance of non-locality is again not genuine, however. What have changed as a result of Alice's measurement are the relative states of Bob's system; that is, roughly, relational properties of his system. It is no mystery that relational properties can be affected unilaterally by operations on one of the \textit{relata} and it certainly does not connote non-locality. (Compare `$x$ is heavier than $y$'; we might make $y$ heavier by adding weights, so that this statement becomes false, but this would not indicate a non-local effect on $x$.) The effect of Alice's measurement has been to entangle further systems with the initial entangled pair, namely, the system whose state was to be transmitted and systems recording the outcome of the Bell measurement. The trick is that the type of measurement interaction Alice performs has been chosen such that the way in which the systems recording the outcome of her measurement are allowed to become related to Bob's system (in virtue of the initial entanglement) entails that relative to their outcome recording states, Bob's system will have the required states. That is, the genuine change is in fact all on Alice's side\footnote{For a closely related discussion of the locality of teleportation from the point of view of Everett, see \cite{vaidman}.}. 

The upshot of this discursion is that we are indeed justified in saying that there is no non-locality in Everett. Although local unitary operations can have remarkable effects on the global state in the presence of entanglement, these effects are not such as to impute non-locality. There are no effects on local and non-relational properties of separated systems and, we have suggested, effects on relational properties (effects of the sort we have seen vividly demonstrated in teleportation) do not imply non-locality\footnote{In an interesting recent paper, \citet{dh} claim to have provided an especially local account of quantum mechanics despite the presence of entanglement. It is not clear, however, that their account, which involves unitary, no-collapse quantum mechanics with no determinate values added, confers any benefits with respect to locality not already available on either an Everettian or statistical interpretation; see \citet{erpart2}.}.      

Finally, let us note in closing that there remains a particularly interesting question which we have not addressed here. This is the question of how the notion of spacetime, normally construed as a manifold of points associated with definite physical events, is to be understood at a fundamental level in an Everettian framework. A suitable point of departure for consideration of this topic is the analysis of \citet{guido}.

\section{Summary}

In this paper, we have seen something of the relation of entanglement and relativity via the notion of non-locality. Whilst Bell-inequality violation provides a relatively precise notion of non-locality, or rather, two separate notions, neither implies any direct link from the entanglement of quantum states to concerns about relativity. Wavefunction collapse, on the other hand, does raise \textit{prima facie} concerns about Lorentz covariance, and we saw how Einstein used entanglement and collapse in his argument for the incompleteness of quantum mechanics. One might give up collapse, however, whilst maintaining that quantum mechanics is a complete theory, leading to the Everett interpretation. Here we saw how co-existence of measurement outcomes alleviates any fear of non-locality in Bell-type experiments; and it was argued further that within the Everett framework, the potentially worrisome global effects of local unitary operations, demonstrated in entanglement assisted communication protocols, would not amount to a novel sort of non-locality.


\begin{thebibliography}{}

\bibitem[Bacciagaluppi, 2002]{guido}
Bacciagaluppi, G. (2002).
\newblock Remarks on space-time and locality in {E}verett's interpretation.
\newblock In Butterfield, J. and Placek, T., editors, {\em Non-Locality and
  Modality}, volume~64 of {\em NATO Science Series: {II}}. Kluwer Academic.
\newblock {\sc Pitt-Phil-Sci} 00000504.

\bibitem[Barrett, 2001]{jon1}
Barrett, J.~S. (2001).
\newblock Implications of teleportation for nonlocality.
\newblock {\em Phys. Rev. A}, 64:042305.
\newblock {a}r{X}iv:quant-ph/0103105.

\bibitem[Barrett, 2002]{jon2}
Barrett, J.~S. (2002).
\newblock Nonsequential positive operator-valued measurements on entangled
  mixed states do not always violate a {B}ell inequality.
\newblock {\em Phys. Rev. A}, 65:042302.
\newblock {a}r{X}iv:quant-ph/0107045.

\bibitem[Bell, 1964]{bell1964}
Bell, J.~S. (1964).
\newblock On the {E}instein-{P}odolsky-{R}osen paradox.
\newblock {\em Physics}, 1:195--200.

\bibitem[Bell, 1966]{bellhiddenvariable}
Bell, J.~S. (1966).
\newblock On the problem of hidden variables in quantum mechanics.
\newblock {\em Rev. Mod. Phys.}, 38:447--52.
\newblock This paper, written before \citep{bell1964}, actually appeared in
  print after it.

\bibitem[Bell, 1976]{bell1976}
Bell, J.~S. (1976).
\newblock The theory of local beables.
\newblock {\em Epistemological Letters}.
\newblock Reprinted, along with \citep{bellhiddenvariable,bell1964} in Bell,
  J.S. \textit{Speakable and Unspeakable in Quantum Mechanics}, Cambridge
  University Press, 1987.

\bibitem[Bennett et~al., 1993]{teleportation}
Bennett, C.~H., Brassard, G., Cr\'epeau, C., Jozsa, R., Peres, A., and
  Wootters, W. (1993).
\newblock Teleporting an unknown state via dual classical and {EPR} channels.
\newblock {\em Phys. Rev. Lett.}, 70:1895--1899.

\bibitem[Bennett and Weisner, 1992]{superdense}
Bennett, C.~H. and Weisner, S.~J. (1992).
\newblock Communication via one- and two-particle operators on
  {E}instein-{P}odolsky-{R}osen states.
\newblock {\em Phys. Rev. Lett.}, 69(20):2881--2884.

\bibitem[Bohm, 1951]{bohm}
Bohm, D. (1951).
\newblock {\em Quantum Theory}, chapter~22, pages 615--619.
\newblock Prentice-Hall, Englewood Cliffs.

\bibitem[Born et~al., 1971]{borneinstein}
Born, M., Born, H., and Einstein, A. (1971).
\newblock {\em The {B}orn-{E}instein Letters}.
\newblock Macmillan.

\bibitem[Braunstein et~al., 2000]{braunstein:twist}
Braunstein, S.~L., D'{A}riano, G.~M., Milburn, G.~J., and Sacchi, M.~F. (2000).
\newblock Universal teleportation with a twist.
\newblock {\em Phys. Rev. Lett.}, 84(15):3486--3489.

\bibitem[Brown, 1991]{harvey:pass}
Brown, H.~R. (1991).
\newblock Nonlocality in quantum mechanics.
\newblock {\em Proc. Aristot. Soc. Supp.}, {LXV}:141--159.

\bibitem[Brown et~al., 1996]{brownelbyweingard:1996}
Brown, H.~R., Elby, A., and Weingard, R. (1996).
\newblock Cause and effect in the pilot wave interpretation of quantum
  mechanics.
\newblock In Cushing, J., Fine, A., and Goldstein, S., editors, {\em Bohmian
  Mechanics and Quantum Theory: An Appraisal}, pages 309--319. Kluwer Academic
  Publishers.

\bibitem[Clauser and Horne, 1974]{ch1974}
Clauser, J.~F. and Horne, M.~A. (1974).
\newblock Experimental consequences of objective local theories.
\newblock {\em Phys. Rev. D}, 10:526.

\bibitem[Deutsch and Hayden, 2000]{dh}
Deutsch, D. and Hayden, P. (2000).
\newblock Information flow in entangled quantum systems.
\newblock {\em Proc. R. Soc. Lond. A}, 456:1759--1774.
\newblock {a}r{X}iv:quant-ph/9906007.

\bibitem[Dickson, 1998]{dickson}
Dickson, W.~M. (1998).
\newblock {\em Quantum Chance and Non-Locality}.
\newblock Cambridge University Press.

\bibitem[Eberhard, 1978]{eberhard}
Eberhard, P.~H. (1978).
\newblock Bell's theorem and the different concepts of locality.
\newblock {\em Nouv. Cim.}, 46B:392--419.

\bibitem[Einstein, 1905]{Einstein1905}
Einstein, A. (1905).
\newblock On the electrodynamics of moving bodies.
\newblock {\em Ann. Phys.}, 17:891--921.
\newblock Repr. in J. Stachel ed. \textit{Einstein's Miraculous Year},
  Princeton University Press, 1998.

\bibitem[Einstein, 1948]{einstein}
Einstein, A. (1948).
\newblock Quantum mechanics and reality.
\newblock {\em Dialetica}, 2:320--324.
\newblock repr. in \citet{borneinstein} pp. 168-173.

\bibitem[Einstein et~al., 1935]{EPR}
Einstein, A., Podolsky, B., and Rosen, N. (1935).
\newblock Can quantum mechanical description of physical reality be considered
  complete?
\newblock {\em Phys. Rev.}, 47:777.

\bibitem[Ekert and Jozsa, 1998]{ekertjozsa}
Ekert, A. and Jozsa, R. (1998).
\newblock Quantum algorithms: Entanglement enhanced information processing.
\newblock {\em Proc. R. Soc. Lond. A}, 356:1769--1782.
\newblock {a}r{X}iv:quant-ph/9803072.

\bibitem[Everett, 1957]{everett}
Everett, III, H. (1957).
\newblock ``{R}elative state" formulation of quantum mechanics.
\newblock {\em Rev. Mod. Phys.}, 29:454--62.

\bibitem[Fine, 1986]{fine}
Fine, A. (1986).
\newblock {\em The Shaky Game}.
\newblock University of Chicago Press.

\bibitem[Fleming, 1989]{fleming}
Fleming, G.~N. (1989).
\newblock Lorentz invariant state reduction and localization.
\newblock In Fine, A. and Leplin, J., editors, {\em PSA 1988}, volume~2.
  Philosophy of Science Association.

\bibitem[Ghiradi, 2002]{ghiradi}
Ghiradi, G.~C. (2002).
\newblock Local measurements of nonlocal observables and the relativistic
  reduction process.
\newblock {\em Found. Phys.}, 30(9):1337--1385.
\newblock {a}r{X}iv:quant-ph/0003149.

\bibitem[Ghiradi et~al., 1980]{grw}
Ghiradi, G.~C., Rimini, A., and Weber, T. (1980).
\newblock A general argument against superluminal transmission through the
  quantum mechanical measurement process.
\newblock {\em Lett. Nuov. Cim.}, 24(10):293--298.

\bibitem[Gingrich and Adami, 2002]{gingrich}
Gingrich, R.~M. and Adami, C. (2002).
\newblock Quantum entanglement of moving bodies.
\newblock {a}r{X}iv:quant-ph/0205179.

\bibitem[Gisin and Peres, 1992]{peresgisin}
Gisin, N. and Peres, A. (1992).
\newblock Maximal violation of {B}ell's inequality for arbitrarily large spin.
\newblock {\em Phys. Lett. A}, 162:15--17.

\bibitem[Horodecki et~al., 2001]{horodeckis}
Horodecki, M., Horodecki, P., and Horodecki, R. (2001).
\newblock Mixed-state entanglement and quantum communication.
\newblock In {\em Quantum Information: An Introduction to Basic Theoretical
  Concepts and Experiments}. Springer-Verlag.
\newblock {a}r{X}iv:quant-ph/0109124.

\bibitem[Howard, 1997]{howard}
Howard, D. (1997).
\newblock Space-time and separability: Problems of identity and individuation
  in fundamental physics.
\newblock In Cohen, R.~S., Horne, M., and Stachel, J., editors, {\em
  Potentiality, Entanglement and Passion-at-a-distance}, pages 113--141. Kluwer
  Academic Publishers.

\bibitem[Jarrett, 1984]{jarrett}
Jarrett, J.~P. (1984).
\newblock On the physical significance of the locality conditions in the {B}ell
  arguments.
\newblock {\em No\^us}, 18:569--589.

\bibitem[Lucas and Hodgson, 1990]{LucasandHodgson}
Lucas, J.~R. and Hodgson, P.~E. (1990).
\newblock {\em Spacetime and Electromagnetism}.
\newblock Oxford University Press.

\bibitem[Maudlin, 1998]{maudlin}
Maudlin, T. (1998).
\newblock Part and whole in quantum mechanics.
\newblock In Castellani, E., editor, {\em Interpreting Bodies}, pages 46--60.
  Princeton University Press, Princeton, New Jersey.

\bibitem[Maudlin, 2002]{Maudlin:non-loc}
Maudlin, T. (2002).
\newblock {\em Quantum Non-Locality and Relativity}.
\newblock Blackwell Publishers Ltd., Oxford, second edition.

\bibitem[Murdoch, 1987]{murdoch}
Murdoch, D. (1987).
\newblock {\em Niels {B}ohr's Philosophy of Physics}.
\newblock Cambridge University Press.

\bibitem[Page, 1982]{page}
Page, D.~N. (1982).
\newblock The {E}instein-{P}odolsky-{R}osen physical reality is completely
  described by quantum mechanics.
\newblock {\em Phys. Lett. A}, 91(2):57--60.

\bibitem[Peres et~al., 2002]{peresqmentropysr}
Peres, A., Scudo, P.~F., and Terno, D.~R. (2002).
\newblock Quantum entropy and special relativity.
\newblock {\em Phys. Rev. Lett.}, 88:230402.
\newblock {a}r{X}iv:quant-ph/0203033.

\bibitem[Popescu, 1995]{popescu}
Popescu, S. (1995).
\newblock Bell's inequalities and density matrices. {R}evealing `hidden'
  non-locality.
\newblock {\em Phys. Rev. Lett.}, 74:2619--2622.
\newblock {a}r{X}iv:quant-ph/9502005.

\bibitem[Popescu and Rohrlich, 1992]{popescurohrlich}
Popescu, S. and Rohrlich, D. (1992).
\newblock Generic quantum nonlocality.
\newblock {\em Phys. Lett. A}, 166:293--297.

\bibitem[Redhead, 1987]{redhead}
Redhead, M. L.~G. (1987).
\newblock {\em Incompleteness, Non-Locality and Realism}, chapter 4.6.
\newblock Oxford University Press.

\bibitem[Saunders, 1995]{simon1}
Saunders, S. (1995).
\newblock Time, quantum mechanics, and decoherence.
\newblock {\em Synthese}, 102:235--266.

\bibitem[Saunders, 1996a]{simonrelativism}
Saunders, S. (1996a).
\newblock Relativism.
\newblock In Clifton, R., editor, {\em Perspectives on Quantum Reality}, pages
  125--142. Kluwer Academic Publishers.

\bibitem[Saunders, 1996b]{simon2}
Saunders, S. (1996b).
\newblock Time, quantum mechanics, and tense.
\newblock {\em Synthese}, 107:19--53.

\bibitem[Saunders, 1998]{simon3}
Saunders, S. (1998).
\newblock Time, quantum mechanics, and probability.
\newblock {\em Synthese}, 114:373--404.

\bibitem[Schilpp, 1979]{einsteinauto}
Schilpp, P.~A., editor (1979).
\newblock {\em {A}lbert {E}instein: Autobiographical Notes}.
\newblock Open Court Publishing Company, La Salle, Illinois.
\newblock First published in P. A. Shilpp ed., \textit{Albert Einstein:
  Philosopher-Scientist} in \textit{The Library of Living Philosophers}.

\bibitem[Schr{\"o}dinger, 1935]{schrodinger1}
Schr{\"o}dinger, E. (1935).
\newblock Discussion of probability relations between separated systems.
\newblock {\em Proc. Camb. Phil. Soc.}, 31:555--563.

\bibitem[Schr{\"o}dinger, 1936]{schrodinger2}
Schr{\"o}dinger, E. (1936).
\newblock Probability relations between separated systems.
\newblock {\em Proc. Camb. Phil. Soc.}, 32:446--452.

\bibitem[Shimony, 1984]{shimony}
Shimony, A. (1984).
\newblock Controllable and uncontrollable non-locality.
\newblock In Kamefuchi, S., editor, {\em Foundations of Quantum Mechanics in
  the Light of New Technology}, Tokyo. The Physical Society of Japan.
\newblock Repr. in Shimony, A. \textit{Search for a Naturalistic World View},
  Vol. 2, Cambridge University Press, 1993, pp.130-139.

\bibitem[Tausk, 1967]{tausk}
Tausk, K. (1967).
\newblock {\em Measurement in Quantum Mechanics}.
\newblock PhD thesis, University of S\~ao Paulo.
\newblock pp.29-31.

\bibitem[Timpson, 2002]{erpart2}
Timpson, C.~G. (2002).
\newblock Entanglement and relativity: The analysis of {D}eutsch and {H}ayden.

\bibitem[Vaidman, 1994]{vaidman}
Vaidman, L. (1994).
\newblock On the paradoxical aspects of new quantum experiments.
\newblock In Hull, D., Forbes, M., and Burian, R., editors, {\em PSA 1994},
  volume~1. Philosophy of Science Association.

\bibitem[Wallace, 2001]{wallace}
Wallace, D. (2001).
\newblock Worlds in the {E}verett interpretation.
\newblock {a}r{X}iv:quant-ph/0103092. To appear in \emph{Stud. Hist. Phil. Mod.
  Phys.}.

\bibitem[Werner, 1989]{werner}
Werner, R.~F. (1989).
\newblock Quantum states with {E}instein-{P}odolsky-{R}osen correlations
  admitting a hidden-variable model.
\newblock {\em Phys. Rev. A}, 40(8):4277--4281.

\end{thebibliography}

\end{document}